\documentclass[fleqn,usenatbib,useAMS]{mnras}

\usepackage[T1]{fontenc}
\usepackage{ae,aecompl}

\usepackage{graphicx}	
\usepackage{amsmath}	
\usepackage{amssymb}	
\usepackage{bm}

\title[DSA-10]{DSA-10: A Prototype Array for Localizing Fast Radio Bursts}

\author[J. Kocz et al.]{
\parbox{\textwidth}{J. Kocz$^{1}$\thanks{E-mail: jkocz@caltech.edu}, 
V. Ravi$^{1}$, 
M. Catha${^2}$, 
L. D'Addario${^1}$, 
G. Hallinan${^1}$, 
R. Hobbs$^{2}$,
S. Kulkarni${^1}$,
J. Shi${^3}$,
H. Vedantham$^{4,1}$,
S. Weinreb$^{1}$,
and D. Woody${^2}$} \vspace{0.4cm}\\
\parbox{\textwidth}{$^1$ Cahill Center for Astronomy, California Institute of Technology, Pasadena, CA\\
$^2$ Owens Valley Radio Observatory, California Institute of Technology, Big Pine, CA\\
$^3$ Department of Electrical Engineering, California Institute of Technology, Pasadena, CA\\
$^4$ ASTRON, Netherlands Institute for Radio Astronomy, Oude Hoogeveensedijk 4, 7991PD, Dwingeloo, The Netherlands.}
}

\date{Accepted XXX. Received YYY; in original form ZZZ}

\pubyear{2019}

\begin{document}
\label{firstpage}
\pagerange{\pageref{firstpage}--\pageref{lastpage}}
\maketitle

\begin{abstract}

The Deep Synoptic Array 10 dish prototype is an instrument designed to detect and localise fast radio bursts with arcsecond accuracy in real time. Deployed at Owens Valley Radio Observatory, it consists of ten 4.5~m diameter dishes, equipped with a 250~MHz bandwidth dual polarisation receiver, centered at 1.4~GHz. The 20 input signals are digitised and field programmable gate arrays are used to transform the data to the frequency domain and transmit it over ethernet. A series of computer servers buffer both raw data samples and perform a real time search for fast radio bursts on the incoherent sum of all inputs. If a pulse is detected, the raw data surrounding the pulse is written to disk for coherent processing and imaging. 

The prototype system was operational from June 2017 - February 2018 conducting a drift scan search. Giant pulses from the Crab pulsar were used to test the detection and imaging pipelines. The 10-dish prototype system was brought online again in March 2019, and will gradually be replaced with the new DSA-110, a 110-dish system, over the next two years to improve sensitivity and localisation accuracy. 

\end{abstract}

\section{Introduction}



Since the discovery of signals \citep[e.g.,][]{keane2012,thornton2013} similar to that described in \citet{lorimer2007}, fast radio bursts (FRBs) have become an active field in radio astronomy, generating many theories as to their origin. However, at the time of writing, only one repeating FRB \citep{chatterjee2017} has been localised to a host galaxy. Building up a sample of well localised bursts will allow us to address questions surrounding the unknown progenitors of FRBs, and the nature of the ionised medium through which they propagate. 

The Deep Synoptic Array 10 dish prototype (DSA-10) is an instrument specifically designed to localise  FRB signals with arcsecond accuracy. This demonstrator project is intended to have a short life cycle, using off the shelf parts for construction as much as possible, and reusing existing infrastructure at the Owens Valley Radio Observatory (OVRO, near Bishop, California). The primary goals of the project are: 1) providing initial constraints on the bright end of the FRB luminosity function, including the possible detection and localisation of a small number of FRBs, and 2) the demonstration of the DSA architecture towards the ongoing 110-dish DSA deployment and the ultimate 2000-dish DSA deployment.

\section{System design}
\label{sec:sysDesign}

The overarching goal of the DSA-10 design was to rapidly deploy an interferometer capable of blindly detecting ultrabright FRBs \citep{lorimer2007,rsb+16,bannister2017}, and localising them to $<\pm2.5$\arcsec~accuracy upon the first instance of detection. The secondary goal was to demonstrate key aspects of the hardware required for future, larger implementations of the DSA concept. The available budget constrained the receiver and digital backend system to service a maximum of ten antennas, with a bandwidth of 250\,MHz at $\sim1.4$\,GHz. Having also settled on using off-the-shelf dishes as the most easily available antenna option, the remaining fundamental design choices were the dish diameter, the center frequency, and the array configuration. 

\subsection{The antennas}

\begin{figure}
	\begin{center}
		\includegraphics[width=0.5\textwidth]{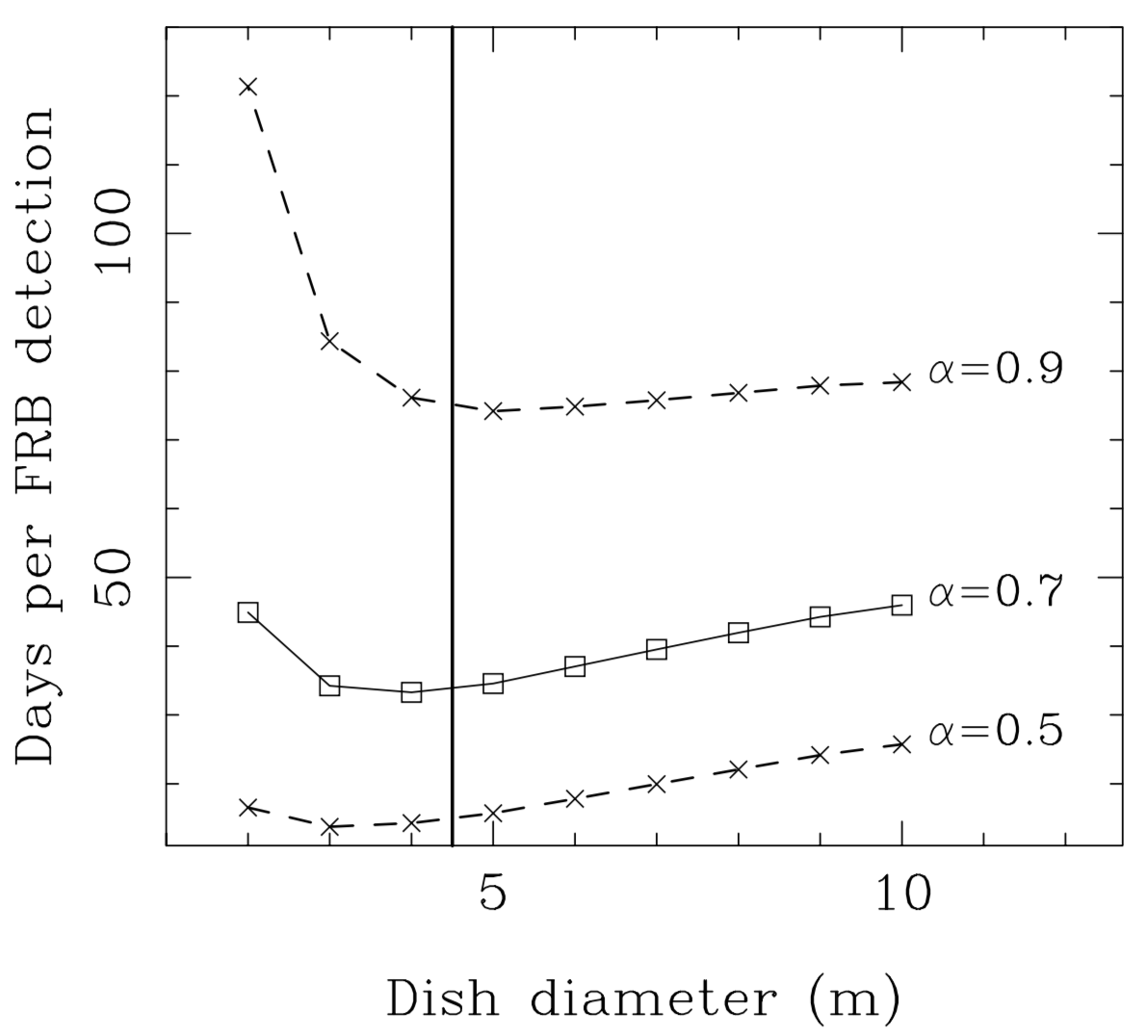}
		\caption{The expected number of days per FRB detection with the DSA-10, for different dish diameters. The \citet{bkb+18} FRB rate was assumed (1700 events per sky per day at fluences $F>2$\,Jy\,ms), and a cumulative fluence distribution of the form $N(>F)\propto F^{-\alpha}$ considered with values of $\alpha=0.5,0.7,0.9$ corresponding to the analysis of \citet{vedantham2016}. The maximum FRB fluence was assumed to be 500\,Jy\,ms, approximately corresponding to the Lorimer burst \citep{r17}. To match the measured DSA-10 performance, a system temperature $T_{\rm sys}=60$\,K, an aperture efficiency $\eta=0.65$, 220\,MHz of useful bandwidth centred on 1405\,MHz, 122.0703125\,kHz channels, 131.072\,$\mu$s sampling, two polarisations, and incoherent summation of all antenna signals for FRB detection are assumed. Finally, empirical distributions for FRB dispersion measures and scattering characteristics were considered marginal, and not included in the model. The vertical line indicates the chosen dish diameter of 4.5\,m.\label{fig:detections}}
	\end{center}
\end{figure}

For a single-beam receiver of a fixed sensitivity mounted on a dish, the FRB detection rate depends primarily on the dish diameter and the fluence distribution of the FRB population \citep[the so-called `logN-logF' curve; e.g.,][]{vedantham2016}. In optimising the DSA-10 dish diameter, the then-recent analysis of \citet{vedantham2016}, who found a shallow fluence distribution such that field of view is likely more important than sensitivity in determining the dish diameter, was assumed. The expected FRB detection rates for different dish diameters are shown in figure~\ref{fig:detections}, which demonstrates how the chosen dish diameter of 4.5\,m was decided. The notion of a shallow logN-logF has been disputed by, e.g., \citet{me18}. However, the baseline sensitivity of the system, discussed below, suggested a detection threshold of 51\,Jy\,ms for a 1\,ms FRB, providing sensitivity to the two brightest FRBs detected at Parkes \citep[FRBs 010124 and 150807;][]{rsb+16,r17} regardless of the fluence distribution. 

\begin{figure}
\begin{center}
\includegraphics[width=6cm]{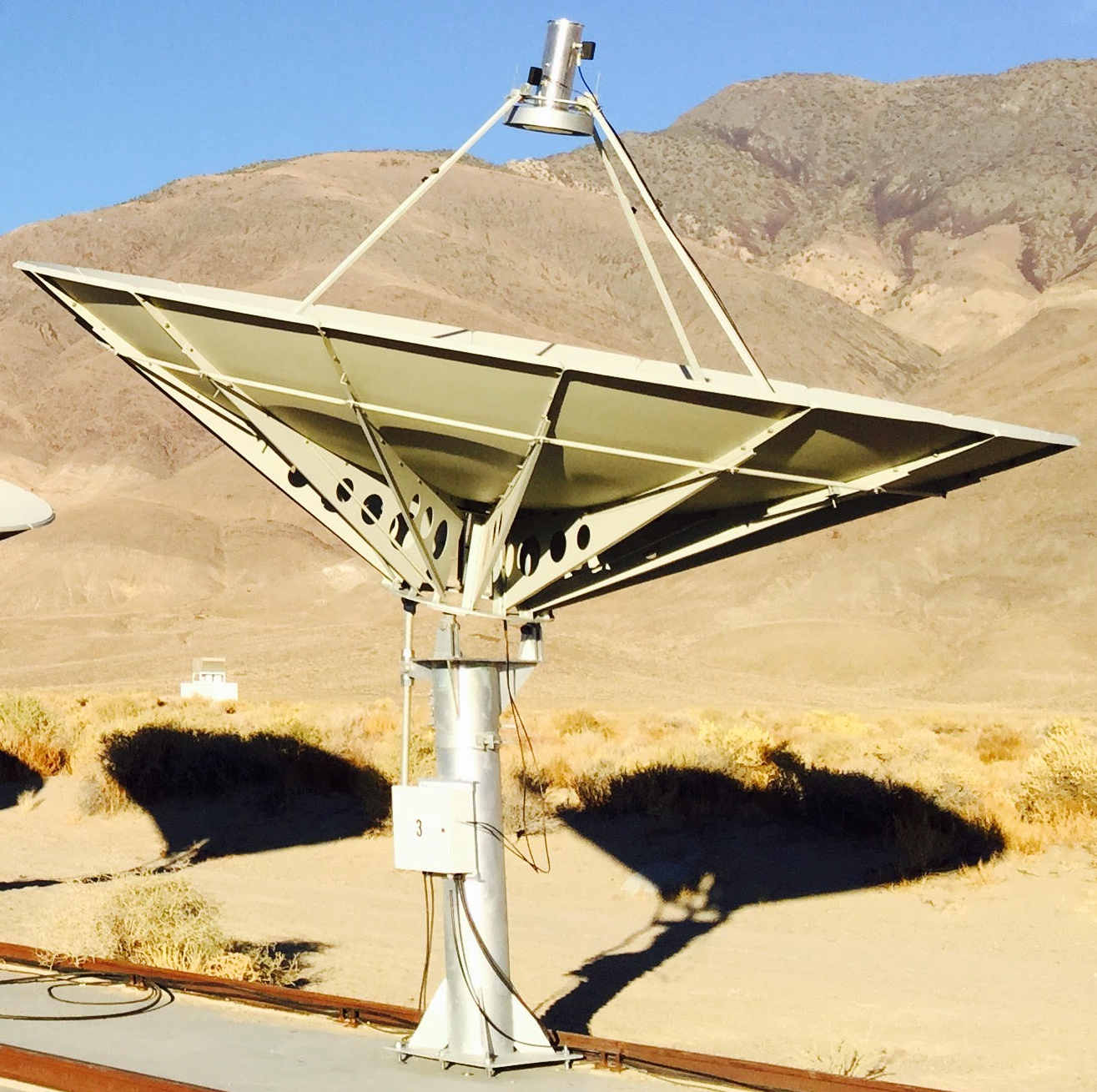}
\caption{An example of an assembled DSA-10 dish, with the receiver mounted. The dish is 4.5~m in diameter, formed of 16 separate aluminium panels, with a surface accuracy of $<0.5$~mm. \label{fig:dsa10}}
\end{center}
\end{figure}

The dishes and mounts were purchased from Hebei Boshida Antenna Equipment (Hebei, China; 4.5\,m C/Ku TVRO antenna). The parabolic dish surfaces were formed from 16 powder-coated solid aluminium panels (surface accuracy $<0.5$\,mm), with steel ribs attached to a central hub as a backing structure. The dishes are mounted on a bearing atop a single central steel post, providing azimuth adjustment, while elevation adjustment is made possible by a manually adjustable jack screw. Four feed support legs are attached to the dish, enabling a prime-focus feed to be mounted 1.7325\,m from the dish vertex (figure \ref{fig:dsa10}). On-sky measurements verified a pointing accuracy of better than 0.4\,deg in all cases. 

\subsection{The feeds and receivers}

\begin{figure*}
	\begin{center}
		\includegraphics[width=18cm]{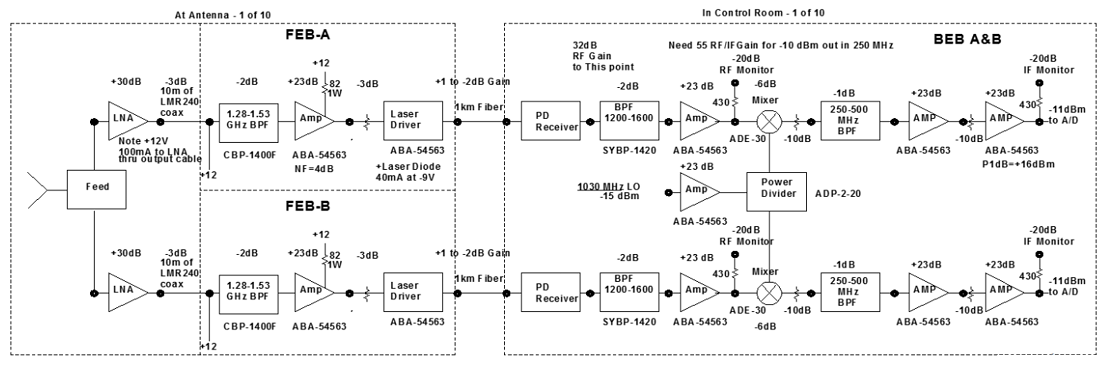}
		\caption{Analogue system receiver chain. The are four main areas of the analogue system. (1) Feed, a dual polarisation 1.28-1.53~GHz feed. (2) LNA, 1300-1700~MHz off the shelf LNA providing 30~dB of gain. (3) FEB, a custom filter, amplifier and optical conversion box for transmitting data back to a central control room. (4) BEB, a custom design to return the optical signal back to RF, filter and down-convert to the 250-500~MHz IF presented to the digital system. \label{fig:arx}}
	\end{center}
\end{figure*}

The DSA-10 analogue system consists of four assemblies for each antenna: a dual polarisation feed, a low noise amplifier (LNA) for each polarisation, a front-end transmitter box (FEB) for each polarisation, and a back-end receiver box (BEB) capable of processing two inputs. The overall receiver chain is shown in figure \ref{fig:arx}. A frequency band of 1.28--1.53\,GHz was chosen based on a survey of the radio frequency interference (RFI) environment of the site. 

\textbf{Feeds.} The dual polarisation feed consists of an aluminum pipe of 5.75'' inner (6'' outer) diameter, 19.1'' long with a 6.5'' diameter end cap, and 15'' mounting disk, situated 3.12'' from the top end of the feed. The two outer concentric rings are 8'' and 15'' in diameter respectively. In the 1.28 to 1.53~GHz range, the feed taper at the edge of an F/D=0.4 dish is 10.7 to 14.9~dB for both polarisations. The on-axis cross-polarisation level is -16 to -20~dB. The resulting full-width half-maximum of the antenna primary beam was found to be 3.25\,deg at the centre of the DSA-10 frequency band.

\textbf{LNAs.} A low-cost, moderate noise LNA often used for amateur radio moon-bounce experiments was implemented in the array. This is the SBA1300/1700\footnote{available from www.g8fek.com} with noise temperature of 32--38\,K and gain of 31$\pm$1\,dB in the 1.28 to 1.53~GHz range. This unit is packaged for outdoor use and has a built-in 1100 MHz high pass filter which gives 50 dB rejection at 800 MHz. The system noise temperature including feed loss, blockage, spillover, sky noise, and contribution of follow-on electronics was between 60--65\,K. 

\textbf{FEBs and RF over fiber links.} The custom designed FEBs, sited within weatherproof boxes attached to each dish mounting post, transmit bandpass-filtered radio-frequency (RF) signals over single-mode fiber to a central correlator room. Each FEB takes the input from a single polarisation, filtering and amplifying the signal, before converting the RF input to a laser signal for transmission over fiber to the control room housing the digital electronics. 

\textbf{BEBs.} The BEBs, also custom designed, are situated in a central control room. Each BEB takes two inputs, converting the modulated laser signals back to RF, before down-converting to an intermediate-frequency (IF) band of 250-500~MHz to be presented to the digital system. 

\subsection{ADC and FPGA processing}

\begin{figure}
	\begin{center}
		\includegraphics[width=7cm]{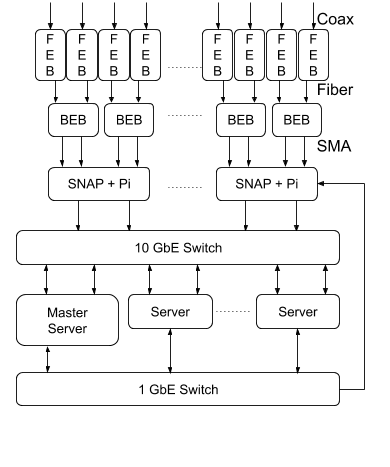}
		\caption{System dataflow overview. A series of ADCs digitise the signal from the BEB before FPGAs covert the signal to the frequency domain using a PFB. The data is then sent via a switch over 10Gb links to a series of servers, which buffer the appropriate raw data, while the incoherent data is sent to a signal node for searching. The system is controlled via 1~Gb ethernet.\label{fig:drx}}
	\end{center}
\end{figure}

\begin{figure*}
	\begin{center}
		\includegraphics[width=14cm]{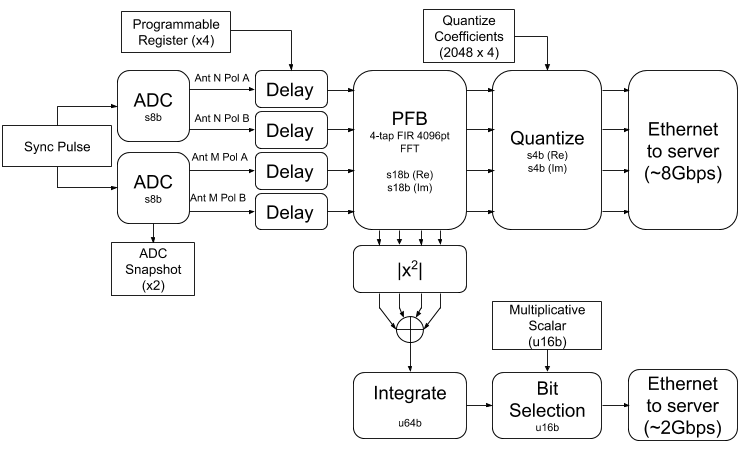}
		\caption{FPGA internals. As the inputs from the BEB are split over multiple SNAPs, each containing ADCs and FPGAs, the system is synchronized via a 1~PPS. A programmable delay block is available for each input to coarsely remove the delays between the different antennas. After being transformed to the frequency domain via PFB, an incoherent sum of each input to the SNAP is made for transmission to a central server. The non-integrated data stream is re-quantized and transmitted at full rate to a separate set of servers (one per SNAP) for buffering, to be written to disk if a candidate is detected in the incoherent sum\label{fig:fpga}}
	\end{center}
\end{figure*}

The analogue signals presented by the BEBs are digitised and packetised using five Smart Network ADC Processor (SNAP)\footnote{https://casper.berkeley.edu/wiki/SNAP} boards. Each board consists of three HMCAD1511 8-bit ADCs (of which two are used), capable of sampling two inputs each at 500~MHz for a total of four signals per board. The ADCs are connected to a Kintex-7 160T FGPA, with an associated dual 10GbE port. Each SNAP board is controlled via Raspberry Pi, which interacts with a controlling PC over Ethernet via a series of python scripts, based on the \texttt{KATCP}\footnote{https://casper.berkeley.edu/wiki/KATCP} protocol. The firmware for the hardware interfaces for the SNAP board (Raspberry Pi, ADCs and 10GbE) is provided by the updated JASPER fork of the CASPER \citep{parsons2008} toolflow. Figure \ref{fig:drx} shows an overview of the dataflow, and figure \ref{fig:fpga} the processing stages in the FPGA. 

Each of the five SNAP boards are provided with a common clock and pulse per second (PPS). This allows synchronisation of the data among the boards. When an observation is to be started, each board is told to start recording on the rising edge of the following PPS. The ADCs are clocked at 500~MHz for a 250~MHz bandwidth. The clock for the FPGA is input via the ADC card. To keep the FPGA fabric running at a reasonably slow speed, the clock and data are demultiplexed by two, with two data samples received from each ADC input on every clock cycle. Each ADC also has a digital gain function, which was used to help equalize the input power from each antenna. 

Once the input signals are digitized, a variable coarse delay correction is applied. This is accomplished by reading each data stream into a BRAM buffer on the FPGA, and waiting for a user defined number of clock cycles before reading the data back out. This gives the course delay a resolution of 4~ns, and allows for the general compensation of cable and other system delays from each of the antennas so that signal coherence can be maintained. 

Once aligned, the data are transformed into the frequency domain via a polyphase filterbank (PFB), implemented via the CASPER libraries using a finite impulse response (FIR) filter and fast Fourier transform (FFT) architecture. The FIR is 4 taps and 4096 points (for 2048 spectral channels) and a Hamming window smoothing function has been applied to the filter coefficients. The filter coefficients are 18-bits, and the 8-bit ADC data are allowed to increase to a maximum of 18 bits real and 18 bits imaginary throughout the PFB processing.

After the PFB, the data path splits into two streams: ``raw'' and ``integrated''. In the raw stream, each 18+18 bit sample is requantized to 4+4 bits. In order to coherently search the data for FRBs, the raw data input from each antenna is required. However, if the 8-bit ADC data were simply transmitted, this would result in a data rate of 8~Gbps per dual pol antenna. By requantizing the data after the PFB, the full 8-bit resolution of the ADC can be used to limit the effect of RFI on the band, while at the same time reducing the overall data rate back to 4~Gbps per dual pol antenna. In order to ensure the minimal amount of data is lost during requantization, it is performed in several steps. For each antenna input, there is a set of 2048 coefficients, corresponding to each spectral channel. The 18+18 bit data are first multiplied by this coefficient, then converted to 4-bits using a round-to-even scheme. If the data exceeds the value that can be represented by 4-bits (positive or negative), it is set to the maximum (or minimum) value. A flag is set in an FPGA register if these saturation values are reached. The requantized raw data, conditioned to have rms values of unity in each channel \citep{ja98}, are transmitted via a 10~Gb switch to five servers, one corresponding to each SNAP. 

In the integrated data stream, the output from the PFB for each antenna and polarisation is squared and combined into a single data stream. As only the incoherent sum from all antennas needs to searched for FRBs (Section \ref{ss:Software}), combining the antennas on each FPGA reduces the data transport load on the system. The resulting spectra on each FPGA are integrated for 16 samples, for a time resolution of 131.072$\mu s$. During integration, the data are allowed to grow to 64 bits resolution. In order to keep the data rate manageable, each integrated spectrum is then multiplied by a single coefficient, and a fixed set of 16-bits are selected from the output (1-16, 17-32, 33-48, or 49-64). Coupled with the incoherent addition of the different antenna inputs, this bit selection reduces the data transmitted for the integrated data stream from 8~Gbps per SNAP board to 2~Gbps. The integrated data are transmitted via a 10~GbE switch to a central node, which further combines the spectra from each of the five SNAPs to a single spectrum for processing by FRB searching software. The specific hardware used is given in table \ref{tab:techSpecs}.

\begin{table}
\begin{tabular}{ c|l|l } 
 Qty & Hardware & Model\\
\hline
\hline
10 & 4.5m Dish & Hebei, China  \\
   &           & 4.5m C/Ku TVRO\\
20 & LNAs & SBA1300/1700  \\
10 & Feed/Receiver & Custom by S. Weinreb\\
\hline
5 & SNAP     & 3x Hittite HMCAD1511\\
    &        & 1x Xilinx Kintex7-160T\\
    &        & 2x 10GbE\\
\hline
6 & GPU Server & 2x 8-core 2.1 GHz CPU\\
 &             & 1x Nvidia GTX 1080 GPU\\
 &             & 1x Mellanox ConnectX-3EN\\
 &             & 96 GB RAM\\
1 & 10Gb Switch & CISCO SG350XG-24F\\
1 & 1Gb Switch  & CISCO SG112-24\\
\end{tabular}
\caption{Summary of hardware specifications}
\label{tab:techSpecs}
\end{table}

\subsection{Data capture and real time processing}
\label{ss:Software}

\begin{figure}
	\begin{center}
		\includegraphics[width=7cm]{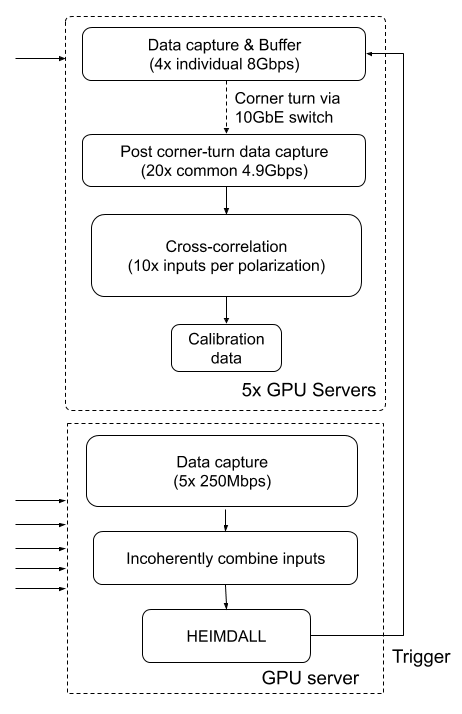}
		\caption{CPU/GPU real time data processing overview. The six computing nodes are partitioned into a single primary, and five secondary nodes. The primary node deals with the incoherently summed and averaged data stream from each SNAP board, combining the data from all boards before using \texttt{HEIMDALL} to search for pulses. The secondary nodes each capture the raw data from an associated SNAP board. This data is buffered to be written to disk in the event of a trigger from the primary node. In addition, each secondary node re-transmits a portion of this data to the other nodes, so each node contains 250 frequency channels from every antenna. A full cross-correlation is performed on this data for calibration purposes.\label{fig:computer}}
	\end{center}
\end{figure}

Data transmitted from the digital back end is captured into memory on the computer side using the \texttt{PSRDADA}\footnote{http://psrdada.sourceforge.net/} software framework. The two real-time data processing streams are outlined below. The software architecture is shown in figure~\ref{fig:computer}. 

\subsubsection{FRB searching}

\begin{figure*}
	\begin{center}
		\includegraphics[width=0.8\textwidth]{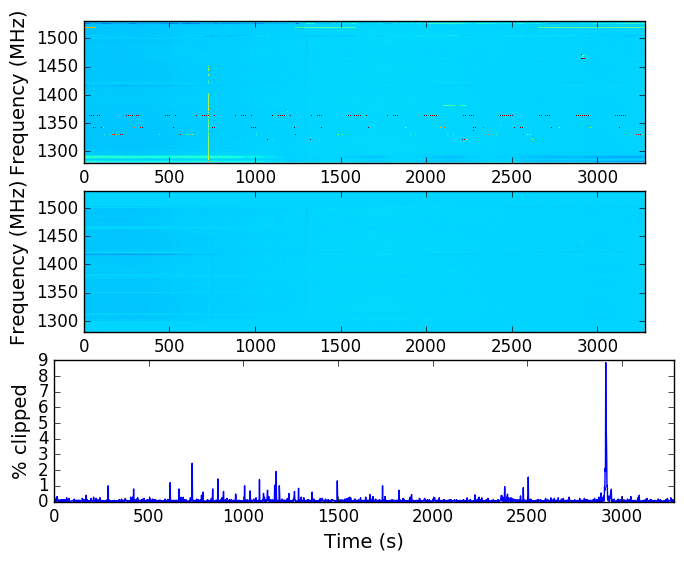}
		\caption{An illustration of the RFI mitigation techniques applied to the incoherently summed data from DSA-10. The top panel shows a series of median-subtracted spectra integrated over 1.6384\,s, prior to any RFI removal. The middle panel shows the same data, on the same colour scale, after removal of RFI in the spectral domain (identified through excess variance or variability in each channel). The bottom panel shows the fraction of data additionally flagged at each time by considering broadband excursions in power. \label{fig:rfi}}
	\end{center}
\end{figure*}

On a designated FRB-search node, the detected, integrated (to $1.31072\times10^{-4}$\,s), and summed (by four inputs from each SNAP) data are captured into five ring buffers. Due to the prevalence of impulsive RFI at the site, including within the 1.28 to 1.53~GHz band of interest, several simple mitigation strategies were included in the real time pipeline. These were implemented predominantly to ensure that RFI does not impact the detection of FRB type signals, and does not create a large number of false candidates that may mask an FRB detection. The implemented strategies include:

\begin{itemize}
\item The initial selection of a 250~MHz frequency band within the potential band of operation that contained the least amount of RFI, with sharp band-defining analogue filters.
\item Comparing each bandpass integration to the previous bandpass, and flagging frequency channels that differ significantly (with an arbitrary threshold established to reduce false positives). 
\item Using the median average deviation to flag channels that have an excessive variance. 
\item Collapsing data in frequency after initial flagging, and masking significant excursions in the time series.
\end{itemize}
All masked data were replaced with the data median. The outcome of these strategies is demonstrated in figure~\ref{fig:rfi}. 

Following the application of these RFI rejection strategies to the summed data from each pair of antennas, the five data streams were bandpass-equalized and further summed to produce a single 16-bit data stream for FRB searching. A single instance of the \texttt{HEIMDALL}\footnote{https://sourceforge.net/p/heimdall-astro/wiki/Home/} \citep{barsdell2012} software package was then used search for single pulses in the dispersion measure (DM) range of 30 - 3000\,pc\,cm$^{-3}$ (with pulse widths up to $2^8$ samples, and a DM tolerance of 1.15). The inbuilt RFI flagging in \texttt{HEIMDALL} was not used, as it slowed processing and was not well suited at the RFI environment. If a pulse was detected by \texttt{HEIMDALL} above a signal to noise (S/N) ratio of 7, and was determined by an automatic inspection of its spectrum to not occupy only a few channels, a trigger was sent to all nodes to write the raw data held in memory to disk. In order to save on both disk space and potential observing down time, only the data surrounding the pulse are written. The SNR threshold is a run time changeable parameter, and is adjusted to have a balance between writing data surrounding the potential detection to disk, and minimising disk write so that the buffer is able to be cleared as new data arrives.

\subsubsection{Calibration and FRB localisation}

The unintegrated, 4-bit real / 4-bit imaginary spectra from each input were captured in \texttt{PSRDADA} ring buffers on each of five secondary nodes. These data were buffered for the reception of candidate triggers from the FRB-search node. In addition, a partial corner turn of 1250 out of 2048 frequency channels was performed on each node using the 10Gb switch. This was accomplished by creating several small memory buffers (one for each node), and transferring the data via custom UDP-based code. The data transfer was arranged such that each node ended up with a subset of frequency channels for every antenna (250 frequency channels, approximately 30~MHz bandwidth). Each of the nodes then performs cross-correlation for all antennas across this frequency range using the \texttt{xGPU} software package \citep{clark2012}. This allows for simple bandpass and phase calibration (described in Section \ref{sec:calibration}), critical for the later imaging steps (figure \ref{fig:computer}).

\subsection{Antenna layout}
\label{ss:layout}

\begin{figure}
	\begin{center}
		\includegraphics[width=6cm]{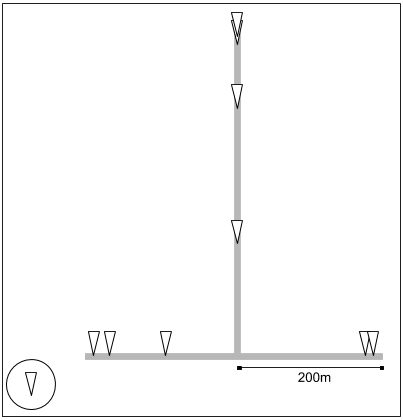}
		\caption{DSA-10 antenna locations. Nine of the antennas are placed on the pre-existing ``T'' (Tee) shaped infrastructure at the observatory, an $\approx 430$ (NS) x 400 (EW) layout, with a tenth (circled) located to the west, providing a long baseline to aid in localisation. The figure is to scale, with the exception of antenna 10, which is located 990~m west, and 217~m south of the center of the Tee.\label{fig:layout}}
	\end{center}
\end{figure}

\begin{figure}
    \centering
    \includegraphics[width=0.49\textwidth]{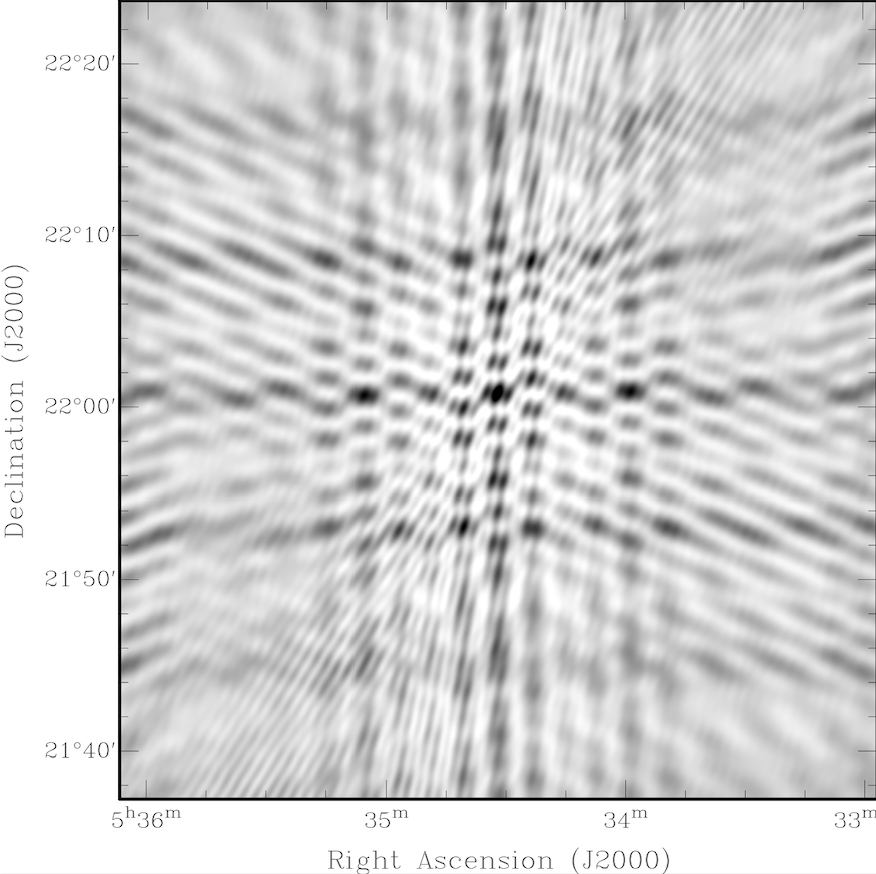}
    \caption{Simulated interferometric point-spread function of DSA-10 at the declination of the Crab pulsar.}
    \label{fig:beam}
\end{figure}

The layout for the ten 4.5~m dishes was determined by optimising for the lowest ambiguity in the synthesised beam. The worst fringe ambiguity was expected between the main lobe of the synthesised beam and the immediately adjacent sidelobes. For environmental reasons, it was required that the antennas be placed in locations previously prepared for antenna use at OVRO. This limited the majority of the placement to the Tee-shaped 442~m by 400~m stretch of previously prepared antenna infrastructure. Additional locations for single antennas were also available near other active antennas (27~m and 40~m) on site. The antenna locations were determined by optimising the sidelobe level assuming a broadband signal (200\,MHz), and accounting for the synthesis of multi-frequency data. The resulting positions are shown in figure \ref{fig:layout}, with the peak sidelobe level of the synthesised beam found to be 0.53 for a broadband burst (figure \ref{fig:beam}). At the declination of the Crab pulsar (approx. $+22$\,deg), this layout corresponds to a synthesised beam shape of $65\times32$\,arcsec (full-width half-maximum). 

\subsection{Interferometric calibration and imaging}
\label{sec:calibration}

A custom suite of software was developed to enable the coherent combination of data from DSA-10. This software was required to fulfill the following tasks:
\begin{enumerate}
\item Process the visibilities on the 45 baselines of the array to derive the per-receiver complex, frequency dependent gains.
\item Process the buffered outputs of the PFBs for a candidate astrophysical pulse to attempt to identify and localise the pulse in an interferometric image. 
\end{enumerate}
A custom software solution was combined with existing, publicly available implementations for radio interferometry because of the unique requirements of DSA-10. These included (a) the operation of DSA-10 as a transit instrument with no online correction for time-variable instrumental delays towards a specific point on the celestial sphere, and (b) the straightforward nature of the imaging problem did not always require the full functionality of existing implementations with regards to calibration and imaging. 

\subsubsection{Calibration}

Consider a baseline between antennas $p$ and $q$ specified by standard coordinates $\mathbf{b_{pq}} = (u_{pq},\,v_{pq},\,w_{pq})$ in units of wavelengths. The visibility measured on this baseline towards an unresolved source with direction cosines $\bm{\sigma} = (l,\,m,\,n)$ defined with respect to the pointing centre is given by 
\begin{equation}
V_{pq}(\nu) = S(\nu)g_{p}(\nu,\bm{\theta})g_{q}(\nu,\bm{\theta})\exp (-2\pi i \mathbf{b_{pq}(\nu)}^{T}\bm{\sigma}) + N.
\end{equation}
Here, $S$ is the flux density of the source, and $g_{p}(\nu,\bm{\theta})$ and $g_{q}(\nu,\bm{\theta})$ are the complex frequency dependent gains for antennas $p$ and $q$ respectively as functions of the observing frequency, $\nu$, and the location of the source in the primary beams of each antenna specified by the angle $\bm{\theta}$. Also, $N$ is a normally distributed complex random variable. The International Terrestrial Reference Frame baseline coordinates, $\mathbf{b_{pq}}$, were determined using \texttt{casacore} routines from antenna positions measured to $\sim10$\,cm accuracy on existing infrastructure at OVRO. 

The gains are solved up to a constant scalar factor by specifying a model for the observed visibilities:
\begin{equation}
M_{pq}(\nu) = \exp (-2\pi i \mathbf{b_{pq}(\nu)}^{T}\bm{\sigma}),
\end{equation}
and using it to derive the quantities
\begin{equation}
E_{pq}(\nu) = \frac{V_{pq}(\nu)}{M_{pq}(\nu)} = S(\nu)g_{p}(\nu,\bm{\theta})g_{q}(\nu,\bm{\theta}) + N',
\end{equation}
where $N'$ is a scaled version of $N$. The coarse antenna-specific delays are determined with respect to a reference antenna by averaging measurements of $E_{pq}(\nu)$ over sufficient time to obtain per-baseline signal to noise ratios of $\sim10$, and Fourier transforming over the frequency axis. These delays, determined to the nearest two ADC samples (i.e., 4\,ns), are corrected online using the SNAP delay registers. After the removal of significant narrow-band and impulsive RFI, the antenna-specific gain terms, $g_{p}(\nu,\bm{\theta})$, are solved up to a constant factor using a global Levenberg-Marquardt minimisation routine applied to all measurements of $E_{pq}(\nu)$. Averaging is typically conducted over 60\,s in time and over 30 frequency channels (3.662\,MHz) to obtain a sufficient signal to noise ratio for the minimization algorithm to effectively converge. Tests on exceedingly bright sources, such as Taurus\,A, Virgo\,A, and the Sun on short baselines, verified that negligible frequency-variation was present in the gains on $<5$\,MHz scales. This enabled the effective calibration of the system despite the lack of a sufficiently bright bandpass calibrator at most declinations. 

In the course of characterising the system, it was found that the phases of the complex antenna gains were not always independent of $\bm{\theta}$, even at the half-power points of the primary beams. This was attributed to inaccuracies in the antenna optics. However, reasonable results could be obtained by using calibrators observed at the same hour angles as sources that needed to be localised where initial guesses at the source hour angles could be obtained using calibration solutions obtained at boresight. Calibration was found to be possible using any unresolved NVSS source with $\gtrsim 2.5$\,Jy of flux apparently present after accounting for off-boresight attenuation. No baseline-dependent calibrations were required. 

\subsubsection{Imaging}

Upon the detection of each candidate pulse, exactly 2.415919104\,s of buffered raw PFB output data (294912 spectra sampled at 8.192\,$\mu$s) were stored to disk for each receiver. The data were aligned such that the pulse arrival time in the highest-frequency channel was 0.5\,s into the raw data, enabling data for pulses with DMs up to 2520\,pc\,cm$^{-3}$ to be fully captured. Correlation products on all 45 baselines were then formed using software identical to the online correlation software, and the antenna-specific calibration solutions (interpolated in frequency) were applied prior to imaging. 

For candidate pulses with unknown locations within the primary beam, images were initially formed between the nulls of the beam response ($7\times7$\,deg), with 15\arcsec~pixels. A standard imaging algorithm was implemented, with nearest-neighbour gridding and no $uv$ taper. No anti-aliasing was necessary because of the high likelihood of any detected pulse being the brightest relevant source either within or beyond the primary beam on the $\sim$ms snapshot timescales. No deconvolution algorithm was implemented for the same reason. Once candidate pulses were identified in images, the code provides functionality to produce zoom-in images with higher resolution, and to also derive time series of the intensity towards a point in the sky by coherently phasing up the visibilities. Additional software was developed to convert data into \texttt{CASA} Measurement Sets for further analysis requirements (e.g., deconvolution, fitting of source positions). 

It should be noted that as initial pulse detection is accomplished by incoherently summing filterbank data from all antennas, pulses can be detected anywhere within the mean primary beam response of the array elements. However, given the high pulse detection threshold ($\sim60$~Jy~ms for a 1~ms pulse), and the ability to subtract off-pulse visibility data, the formed images include just one unresolved point source at the pulse location. The position of this source can then be determined in several ways, such as fitting the synthesised beam shape to the data. Because of this, there is ultimately no ambiguity about whether pulses are detected in the main lobes or sidelobes of either the primary or synthesised beams.

\section{Commissioning}
\label{sec:firstLight}

The DSA-10 was operated in the configuration described here for approximately 6 months on sky during the June 2017 - February 2018 period. No FRBs were detected during that time. Quantitative results of FRB searches with DSA-10 will be presented future publications. 

The sensitivities of individual antennas were initially quantified using the Sun as a reference source, and later using interferometric data. The former measurements revealed system temperatures in the range of 60--65\,K, and aperture efficiencies of approx. 0.6. Between 32--38\,K is contributed by the LNAs, and the remainder is due to spillover ($\sim15$\,K), and a combination of feed losses, RFI and the sky background. These imply typical per-receiver system-equivalent flux densities (SEFDs) of approx. 18\,kJy, and thus a $7\sigma$ detection threshold (assuming 220\,MHz of useful bandwidth) of 60\,Jy\,ms for a 1~ms FRB. 

\begin{figure}
    \centering
    \includegraphics[width=0.49\textwidth]{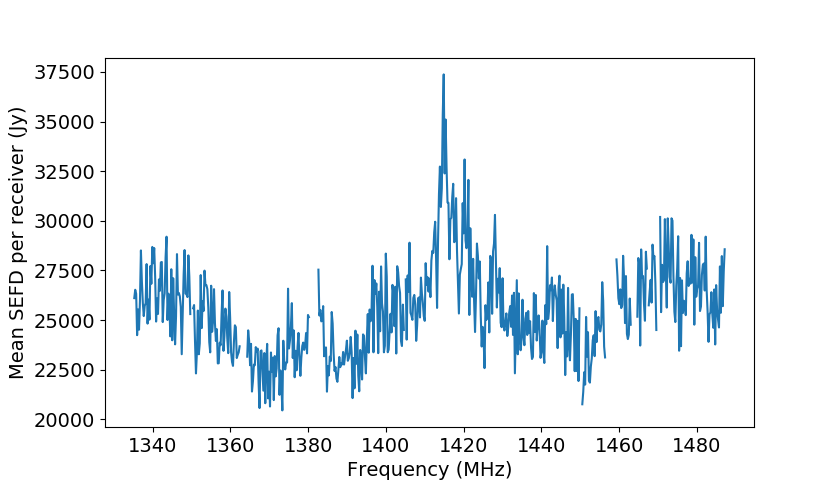}
    \caption{Mean per-antenna system-equivalent flux density (SEFD) estimated using the bright calibrator source 3C409, assumed to have a mean flux density of 13.7\,Jy in the observing band band but observed 1.5\,deg off-boresight. Channels affected by RFI have been excised.}
    \label{fig:sefd}
\end{figure}

An interferometric verification of the system sensitivity is shown in figure~\ref{fig:sefd}. This was based on the signal to noise ratio estimated in visibilities recorded over a five-minute transit of a bright calibrator source 3C409 (assumed to have a mean flux density of 13.7\,Jy in the observing band but observed 1.5\,deg off-boresight). The feature in the middle of the band is due to a trapped mode in the DSA-10 feeds. Other causes of the decrease in sensitivity with respect to those expected from the single-dish measurements are not fully characterised, but include imperfect phase calibration, broadband correlated RFI, and pointing errors.

\begin{figure}
    \centering
    \includegraphics[width=0.49\textwidth]{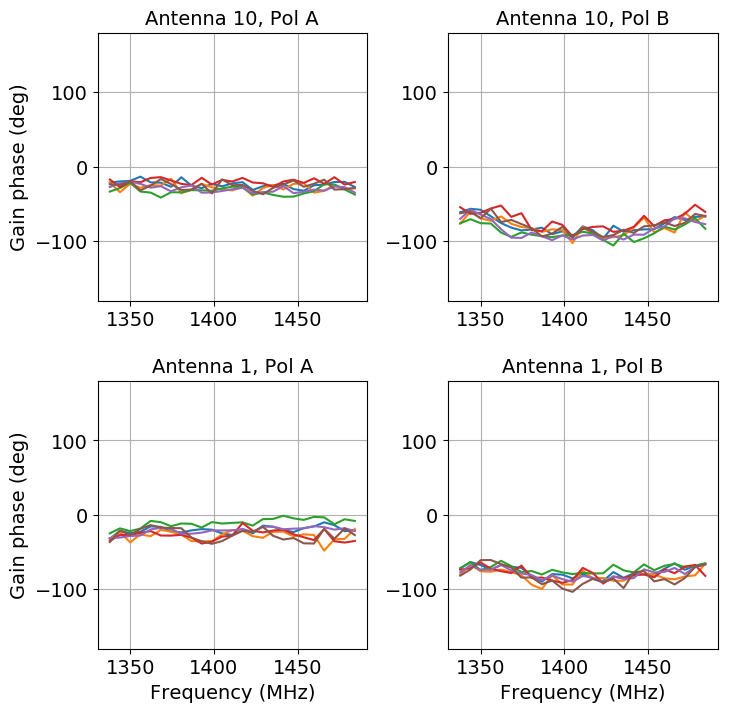}
    \caption{Phases of the complex gains in each polarisation for two antennas estimated using calibrators spaced by approximately 12 hours over three days. Antenna 10 is the sole antenna not sited on the Tee infrastructure, and antenna 1 is the northernmost antenna of the array.}
    \label{fig:gains}
\end{figure}

Another crucial metric of the DSA-10 performance is the stability of the phase-calibration solutions. An example of six consecutive measurements using two calibrators spaced by approx. 12 hours is shown in figure~\ref{fig:gains}. Results are shown for the northernmost antenna on the Tee (Antenna 1), and the sole antenna not located on the Tee (Antenna 10). Little significant variation, either diurnally or over several days, is observed. The typical rms variation of the phase solutions in 3.662\,MHz channels is 10\,deg. 

\begin{figure}
	\begin{center}
		\includegraphics[width=0.49\textwidth]{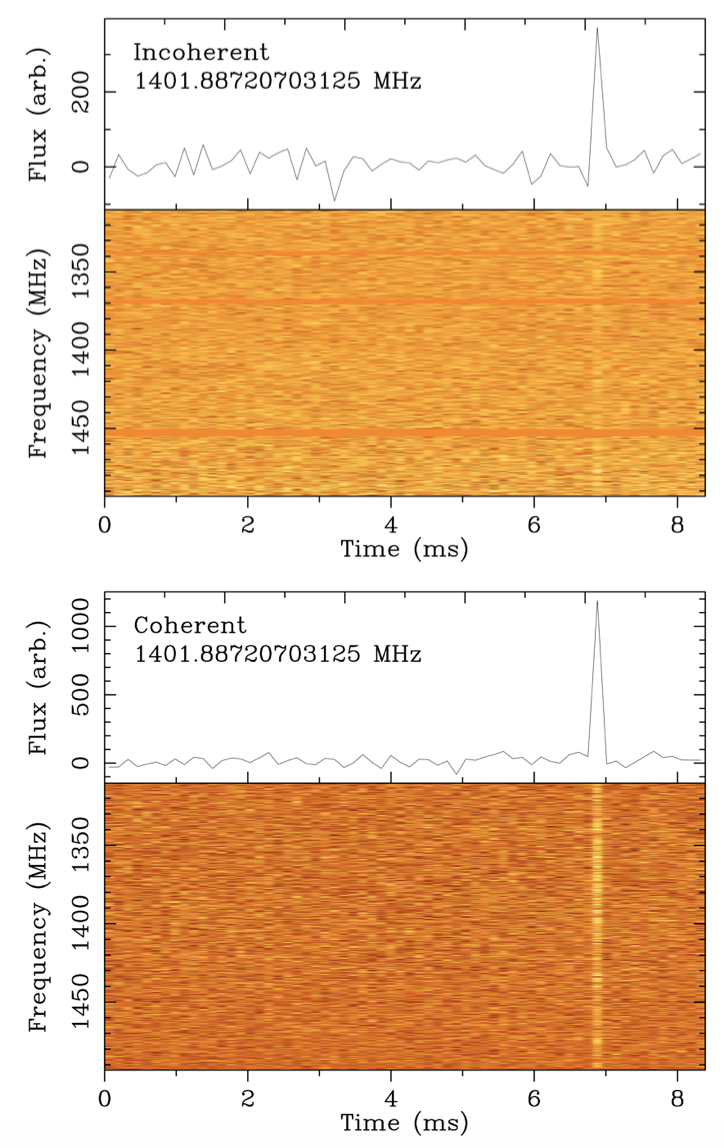}
		\caption{Detection of a giant pulse from the Crab pulsar (B0521$+$21) using DSA-10. Top: Incoherently summed and dedispersed pulse. Bottom: The same pulse, coherently summed and dedispersed. In each figure, the top panel is flux vs time, with flux normalised to the same arbitrary units, showing the summed power across frequency for each time step. The bottom panel of each figure is time vs frequency, showing the dedispersed pulse across the observing band during an approximately 8~ms time window. Channels affected by RFI have been blanked in the top figure. This RFI does not add coherently at the location of the Crab pulsar, and as such they do not affect the bottom figure. \label{fig:crab}}
	\end{center}
\end{figure}

\begin{figure}
	\begin{center}
		\includegraphics[width=0.49\textwidth]{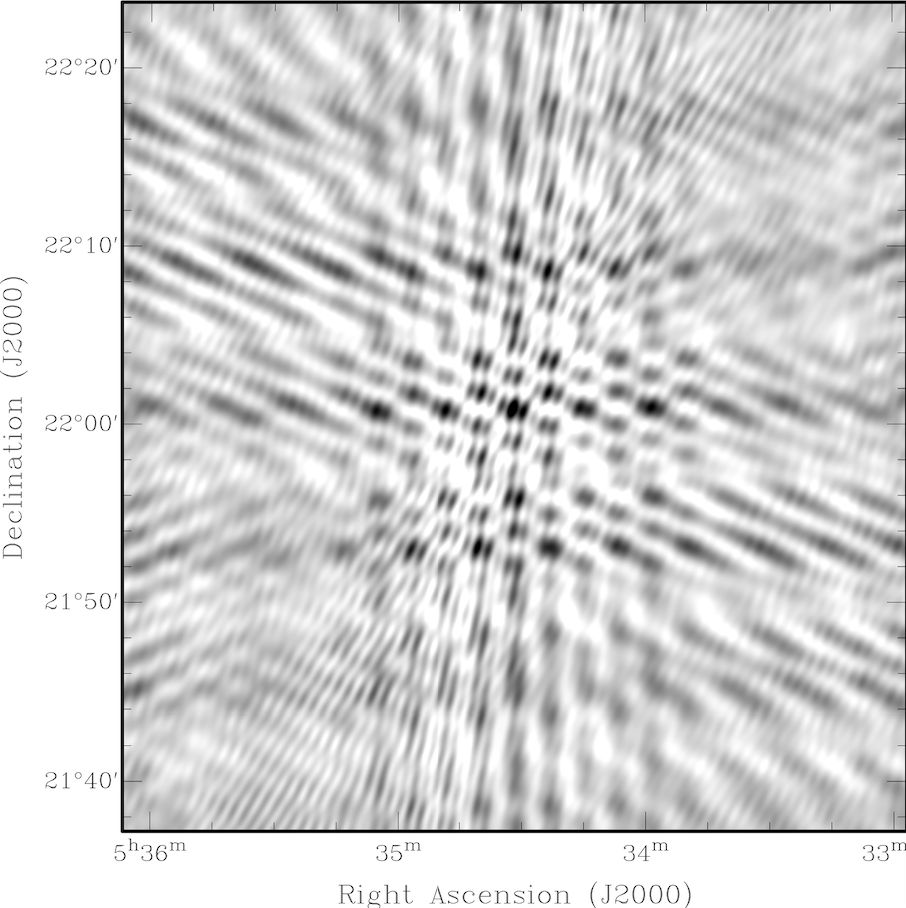}
		\caption{Dirty image obtained using DSA-10 of a giant pulse from the Crab pulsar (B0521$+$21). This compares agreeably with the synthesised beam shape shown in figure~\ref{fig:beam}.\label{fig:dirty}}
	\end{center}
\end{figure}

In order to test the detection pipeline performance, and for continued operational verification, the dishes were aligned so that the Crab pulsar transited the beam. Giant pulses mimic an FRB sufficiently to test all aspects of the pipeline, including: initial detection, writing of raw data, calibration, imaging and localisation. Figure \ref{fig:crab} shows a Crab giant pulse that was detected during transit, with the accompanying PSF (figure \ref{fig:dirty}) which matches the simulated PSF for DSA-10 given in figure \ref{fig:beam}, at the correct coordinates.

\section{Future Observations}
\label{sec:conclustions}

Between June 2017 and February 2018, DSA-10 has been on sky for a total of approximately 6 months, conducting a drift scan searching in real time for FRBs. Giant pulses from the Crab pulsar and daily transits of unresolved continuum sources have been used to test the automated detection, calibration and imaging pipeline, and verified that FRBs with fluences greater than $\approx$ 60~Jy~ms should be detected. Several potential FRB candidates were identified during this observation window, however all were later discarded. The null detection of FRBs over this time period is consistent with an $\alpha > 1$ \citep[e.g.,][]{c16}. 

Beginning in March 2018 the system was offline for testing in various alternative configurations and modes, but was operational again as of March 2019. Over the course of 2019 and 2020, the DSA-10 will be replaced with the DSA-110, an instrument with approximately 100 dishes mounted on the existing Tee, in addition to several outrigger antennas, in order to increase sensitivity and localisation accuracy. The DSA-110 with have 4.75~m dishes with elevation-capable drives and a new custom-designed LNA and feed giving a $T_{\rm sys}$ of approximately 30~K. The digital backend will also be upgraded to operate in a fully coherent mode. DSA-110 is expected to detect and localize over 100 FRBs per year, even the case of an extreme ($>2.5$) logN-logF slope.

\section*{Acknowledgments}

A portion of this research was performed at the Jet Propulsion Laboratory, California Institute of Technology, under a President and Directors Fund grant and under a contract with the National Aeronautics and Space Administration. Copyright 2019 California Institute of Technology. Government sponsorship acknowledged.

\clearpage

\end{document}